\documentclass{appolb}
\usepackage{graphicx}

\begin{document}
\title{Beyond mean-field boson-fermion description of odd nuclei%
\thanks{Presented at XXVI Nuclear Physics Workshop, Kazimierz Dolny, Poland, 24--29 September 2019.}%
}
\author{Kosuke Nomura
\address{
Department of Physics, Faculty of Science, 
University of Zagreb, HR-10000 Zagreb, Croatia}
}
\maketitle

\begin{abstract}
We develop a novel theoretical method for calculating spectroscopic 
properties of those nuclei with odd number of nucleons, 
that is based on the nuclear density functional 
theory and the particle-boson coupling scheme. 
Self-consistent mean-field calculation based on the 
DFT is performed to provide microscopic 
inputs to build the Hamiltonian of the interacting boson-fermion 
systems, which gives excitation spectra and transition rates 
of odd-mass nuclei. 
The method is successfully applied to identify 
the quantum shape phase transitions 
and the role of octupole correlations in odd-mass nuclei, 
and is extended further to odd-odd nuclear systems. 
\end{abstract}

  
\section{Introduction}
Microscopic description of the low-lying states in those nuclei 
with odd numbers of nucleons is one of the challenging 
problems in nuclear structure physics. 
Primary reason is that, as compare to the even-even 
nuclei where nucleons are coupled pairwise and they determine 
to a large extent the low-lying nuclear structure, 
in the odd-A and odd-odd nuclei one has to consider 
both the single-particle (unpaired nucleon) and collective 
degrees of freedom on the same footing. 
Here we specially focus on a recently developed theoretical method 
\cite{nomura2016odd} for calculating spectroscopic properties of the odd nuclei, 
that is based on the nuclear energy 
density functional theory (DFT) 
\cite{bender2003,vretenar2005} and the particle-(boson-)core 
coupling scheme.  
In this framework, the constrained self-consistent 
mean-field (SCMF) calculation based on the nuclear DFT 
is performed to compute 
potential energy surface (PES) in the relevant 
collective coordinates for the even-even core nucleus, 
and the spherical single-particle energies and 
occupation probabilities of odd particles; 
These quantities are used as microscopical input to 
build the interacting boson-fermion 
model (IBFM) \cite{IBFM} Hamiltonian; 
At the cost of having to determine a few coupling 
constants for the boson-fermion interaction 
empirically, this method has allowed 
for a detailed, systematic, and computationally 
feasible description of spectroscopy of those  nuclei with 
odd nucleon numbers. 
Here the recent applications of this method are 
highlighted, i.e., 
the effect of an odd particle on the nature of 
quantum shape phase transitions \cite{nomura2016qpt}, 
the octupole correlations in odd-A neutron-rich 
nuclei \cite{nomura2018oct}, and the extension 
of the method to odd-odd nuclei \cite{nomura2019dodd}.

\section{Theoretical framework\label{sec:model}}

The IBFM for a given 
odd-A system consists of the interacting boson model 
(IBM) \cite{IBM} Hamiltonian $\hat H_\mathrm{B}$ 
for the even-even core nucleus, single-particle 
(either neutron $\rho=\nu$ or proton $\pi$) Hamiltonian 
$\hat H_\mathrm{F}^\rho$, and the interaction 
$\hat H_\mathrm{BF}^\rho$ between 
the odd neutron (proton) and the boson core:
\begin{equation}
\hat H_\mathrm{IBFM} = \hat H_\mathrm{B} + \hat H_\mathrm{F}^\nu 
+ \hat H_\mathrm{F}^\pi  + \hat H_\mathrm{BF}^\nu 
+ \hat H_\mathrm{BF}^\pi. 
\end{equation}
The IBM Hamiltonian is typically of the form: 
\begin{equation}
\hat H_\mathrm{B} = \epsilon(\hat n_{d_\nu}+\hat n_{d_\pi})
+\kappa\hat Q_\nu^{\chi_\nu}\cdot \hat Q_\pi^{\chi_\pi}
+\kappa'\hat L\cdot\hat L,
\end{equation}
where the first and second terms are the $d$-boson 
number operator and quadrupole-quadrupole interaction 
between the neutron and proton bosons, respectively, 
and the third term is the rotational term. 
The strength parameters $\epsilon$, $\kappa$, $\chi_\nu$, 
and $\chi_\pi$ are determined by following the procedure of 
Ref.~\cite{nomura2008}: 
the total mean-field energy obtained from the 
EDF calculation at each $(\beta,\gamma)$ deformation, 
i.e., $E_\mathrm{EDF}(\beta,\gamma)$, 
is equated to the expectation value of the IBM Hamiltonian 
in the intrinsic wave function for the boson system 
at the corresponding configuration, 
$E_\mathrm{IBM}(\beta,\gamma)$. 
Only the parameter $\kappa'$ for the rotational term 
is determined separately from the other parameters, 
in such a way that the SCMF cranking moment of inertia 
at the equilibrium minimum, 
be equal to the IBM counterpart \cite{nomura2011rot}. 

Additional microscopic inputs from the DFT 
are the single-particle energies $\epsilon_{j_\rho}$ 
(for $\hat H_\mathrm{F}^\rho$) and occupation probabilities 
$v^2_{j_\rho}$ (for $\hat H_\mathrm{BF}^\rho$) for the odd nucleon 
at the orbital $j_\rho$. 
These quantities are obtained from the SCMF 
calculation at zero deformation and the particle number of either neutrons 
or protons constrained to odd number \cite{nomura2016odd}. 
The boson-fermion interaction $\hat H_\mathrm{BF}^\rho$ 
is composed of the three essential terms, i.e., 
dynamical quadrupole, exchange, and monopole terms \cite{IBFM}. 
Hence there are three strength parameters 
for $\hat H_\mathrm{BF}^\rho$ for odd-N or 
odd-Z system. 
They are the only phenomenological parameters, and 
are fixed so as to  reasonably reproduce 
the experimental low-lying levels for each odd-A nucleus. 
For the detailed account of the whole procedure, 
the reader is referred to Ref.~\cite{nomura2016odd}.

\section{Signatures of quantum phase transitions in odd-A nuclei}

\begin{figure}[htb]
\centerline{%
\includegraphics[width=8.0cm]{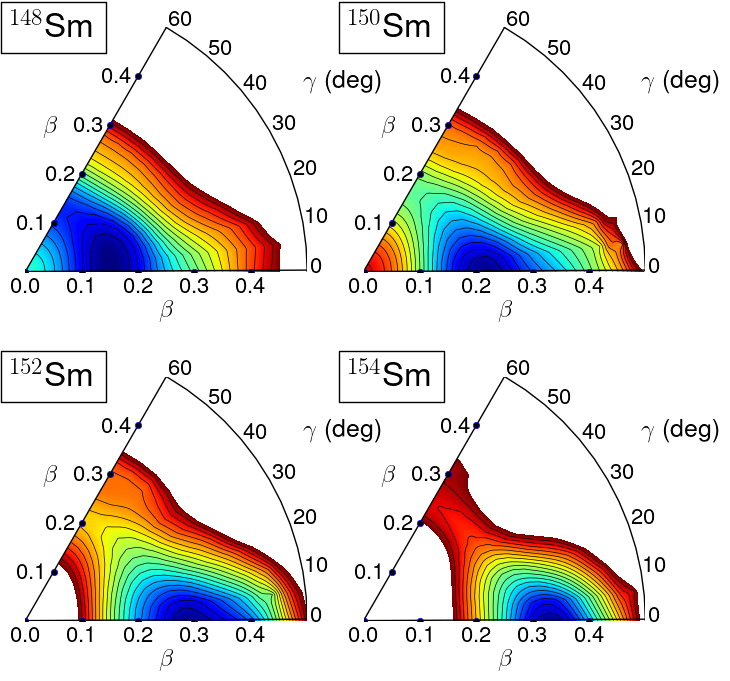}
}
\caption{Potential energy surfaces for the 
Sm isotopes computed with the constrained 
RHB method \cite{vretenar2005} using the 
DD-PC1 EDF \cite{DDPC1}. Energy difference 
between neighboring contours is 250 keV.}
\label{fig:sm}
\end{figure}

\begin{figure}[htb]
\centerline{%
\includegraphics[width=9.0cm]{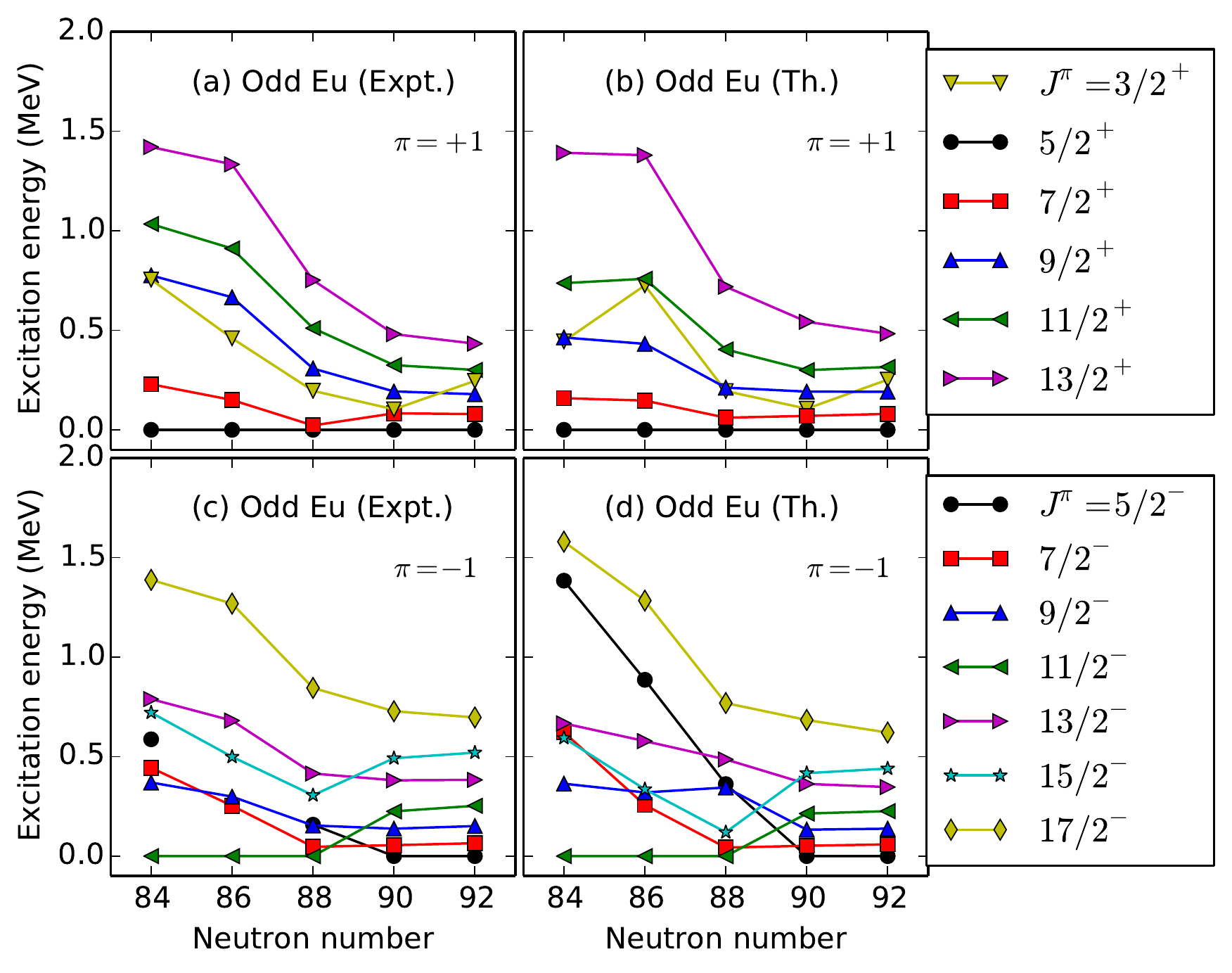} 
}
\centerline{%
\includegraphics[width=9.0cm]{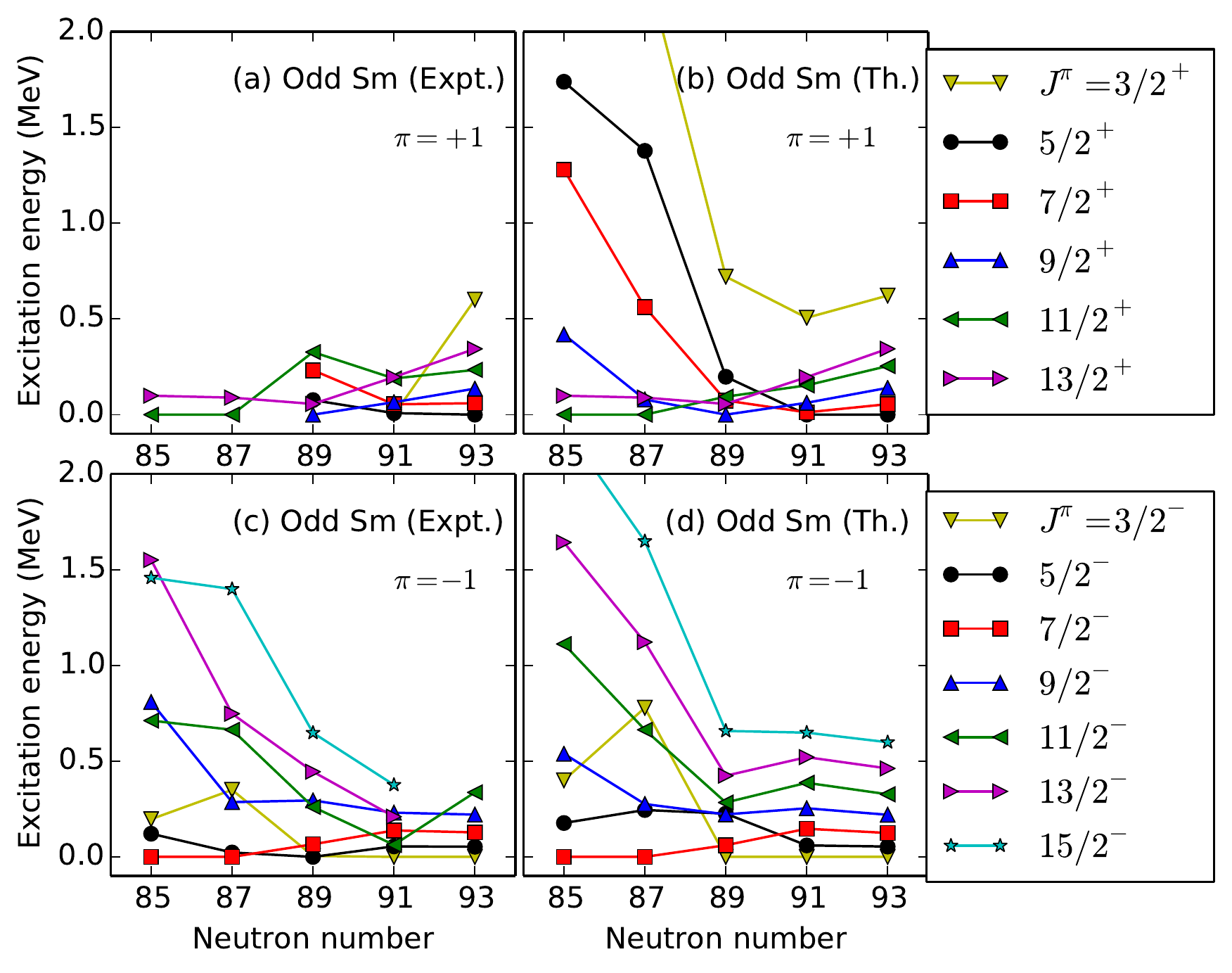}
}
\caption{Excitation spectra for the low-lying positive- 
($\pi=+1$) and negative-parity ($\pi=-1$) states for the odd-A Eu and 
Sm isotopes.}
\label{fig:level}
\end{figure}

\begin{figure}[htb]
\centerline{%
\includegraphics[width=9.0cm]{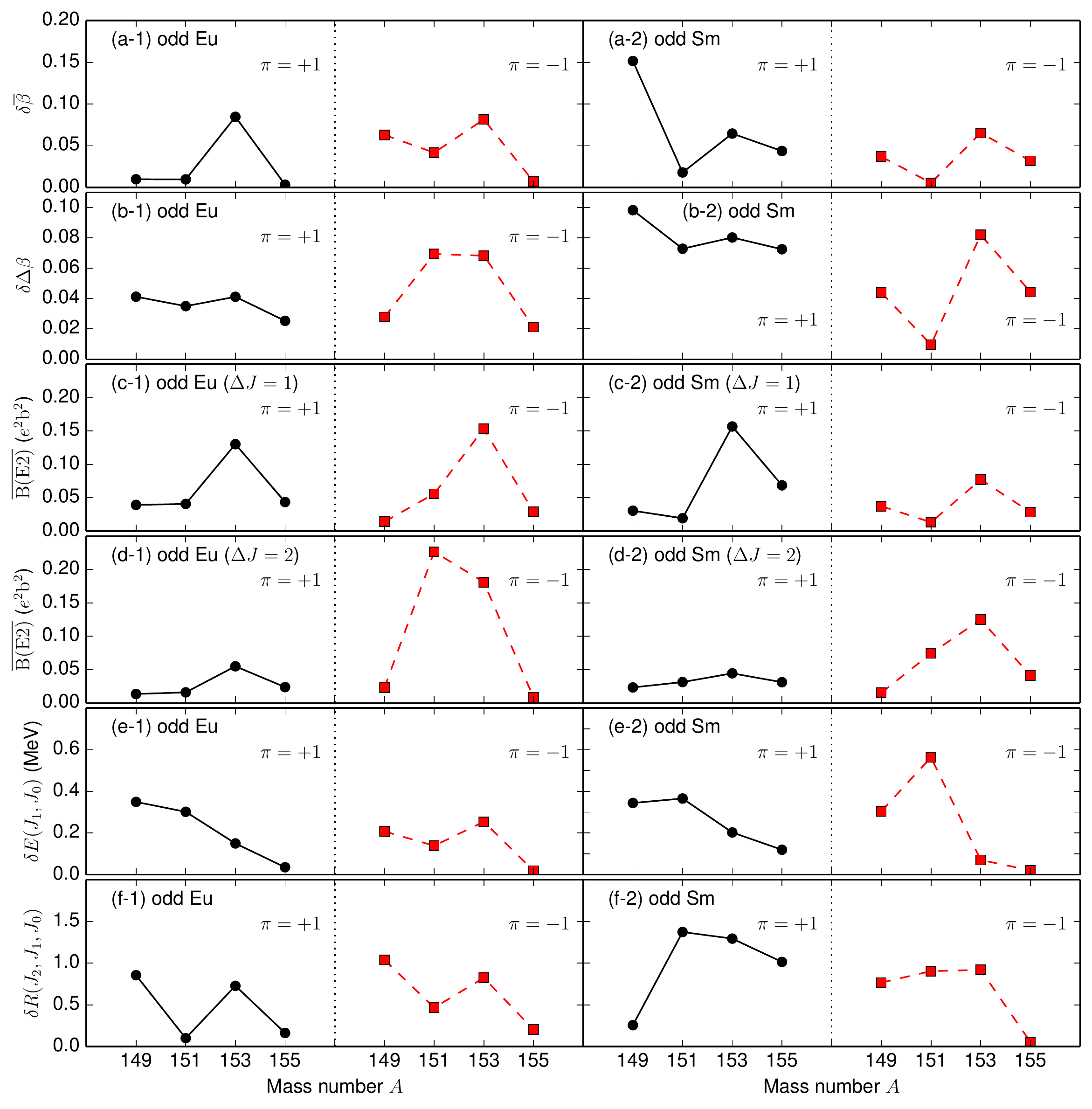} 
}
\caption{Differentials of several calculated 
mean-field and spectroscopic properties of the odd-A Eu and Sm for both parities.}
\label{fig:diff}
\end{figure}

Even-even Sm isotopes are an well-known example 
where the phase transition from spherical vibrational 
to deformed rotational states is suggested to occur 
by addition/subtraction of only a few nucleons \cite{cejnar2010}. 
The triaxial quadrupole $(\beta,\gamma)$ PESs for a 
set of Sm isotopes are depicted in 
Fig.~\ref{fig:sm}, which are obtained by the constrained 
relativistic Hartree-Bogoliubov (RHB) method \cite{vretenar2005} 
with the DD-PC1 EDF \cite{DDPC1}. 
One sees near spherical minimum 
at $^{148}$Sm, which is typical of the vibrational nucleus. 
For $^{150,152}$Sm, the minimum becomes soft 
both in $\beta$ and $\gamma$ deformations, and 
these nuclei are supposed to be the transitional nuclei. 
A distinct prolate minimum is seen in $^{154}$Sm, 
indicating that the deformed rotational structure appears.  
Figure~\ref{fig:level} depicts the excitation spectra for the 
low-lying states of the odd-A $_{63}$Eu and $_{62}$Sm isotopes 
resulting from the IBFM Hamiltonian. Our calculation 
reproduces the experimental spectra quite nicely, 
even though there are only three free parameters. 
A signature of the nuclear structure evolution is identified 
in the change of the ground state spin at $N\approx 90$. 
Around this neutron number, the corresponding even-even 
system is undergoes a rapid shape transition. 
We have further computed several mean-field and 
spectroscopic properties for the odd-A systems, that 
can be considered order parameters for the phase 
transition:  
$\overline{\beta}$ (average $\beta$ deformation), 
$\Delta\beta$ (variance of $\beta$), 
$\overline{B\mathrm{(E2)}}$ (average $B$(E2) value  
between the band-head of a given band with spin $J_0$ 
and the lowest five states with $J_0+\Delta J$, with $\Delta=1,2$), 
$E(J_1,J_0)$ (energy of the first excited state 
$J_1$ in a given band with respect to the 
band-head $J_0$), and 
$R(J_2,J_1,J_0)$ (energy ratio of the second $J_2$ 
to first $J_1$ excited states with respect to the band-head $J_0$ 
in a given band). 
Figure~\ref{fig:diff} depicts the differentials of these 
quantities between the neighboring isotopes, 
$\delta{\cal O}=\sum_i^n|{\cal O}_{i,A}-{\cal O}_{i,(A-2)}|/n$, 
which is averaged over the lowest $n(\approx 5)$ bands. 
One realizes, in most of these calculated quantities, 
a kink around $N\approx 90$, that can be considered 
a signature of phase transition.

\section{Octupole correlations in neutron-rich odd-A nuclei}

\begin{figure}[htb]
\centerline{%
\includegraphics[width=10.0cm]{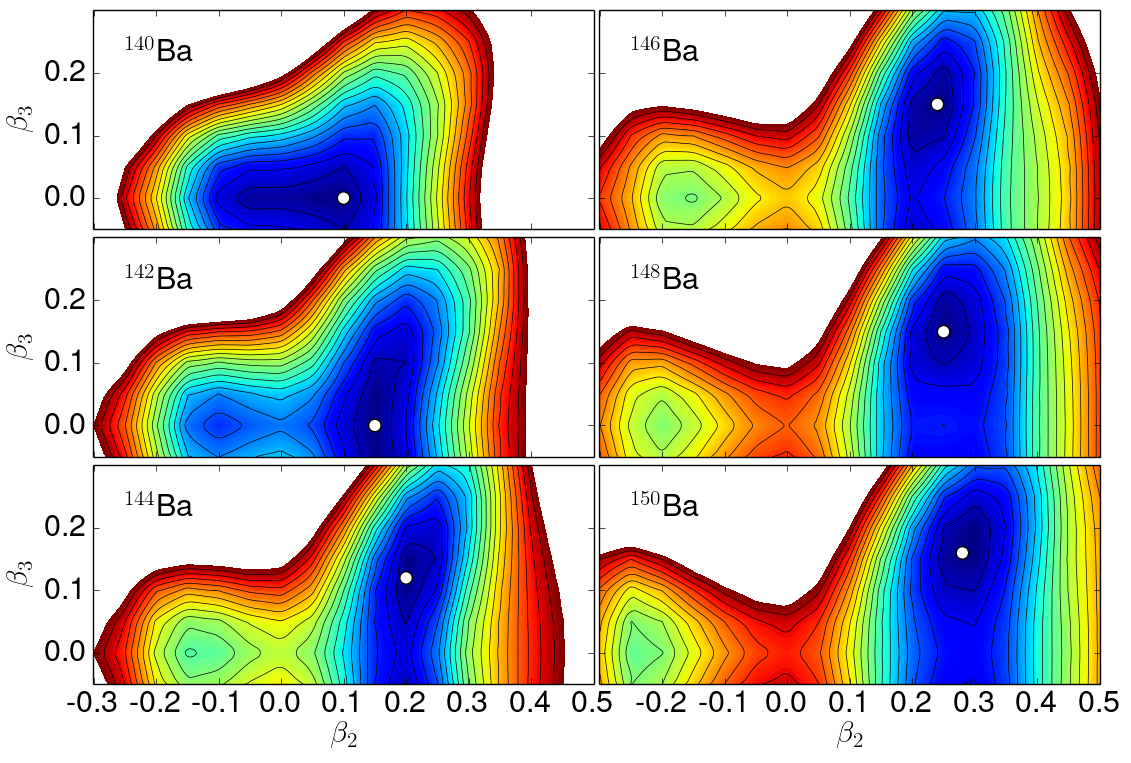}
}
\caption{Axially-symmetric $(\beta_{20},\beta_{30})$ PESs for the 
$^{140-150}$Ba isotopes, computed with the RHB 
method with DD-PC1 EDF.}
\label{fig:oct}
\end{figure}

\begin{figure}[htb]
\centerline{%
\includegraphics[width=6.0cm]{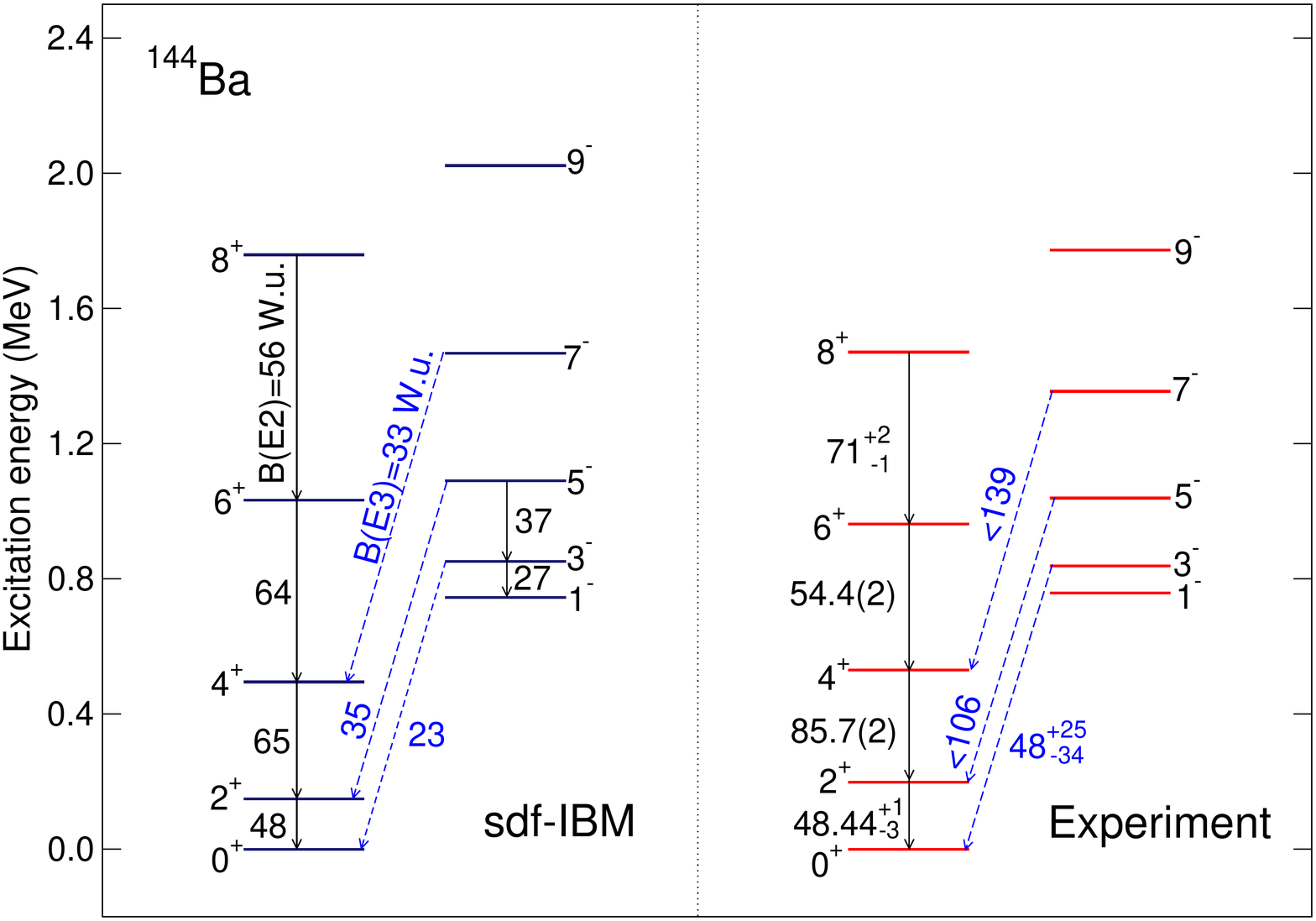}
\includegraphics[width=6.0cm]{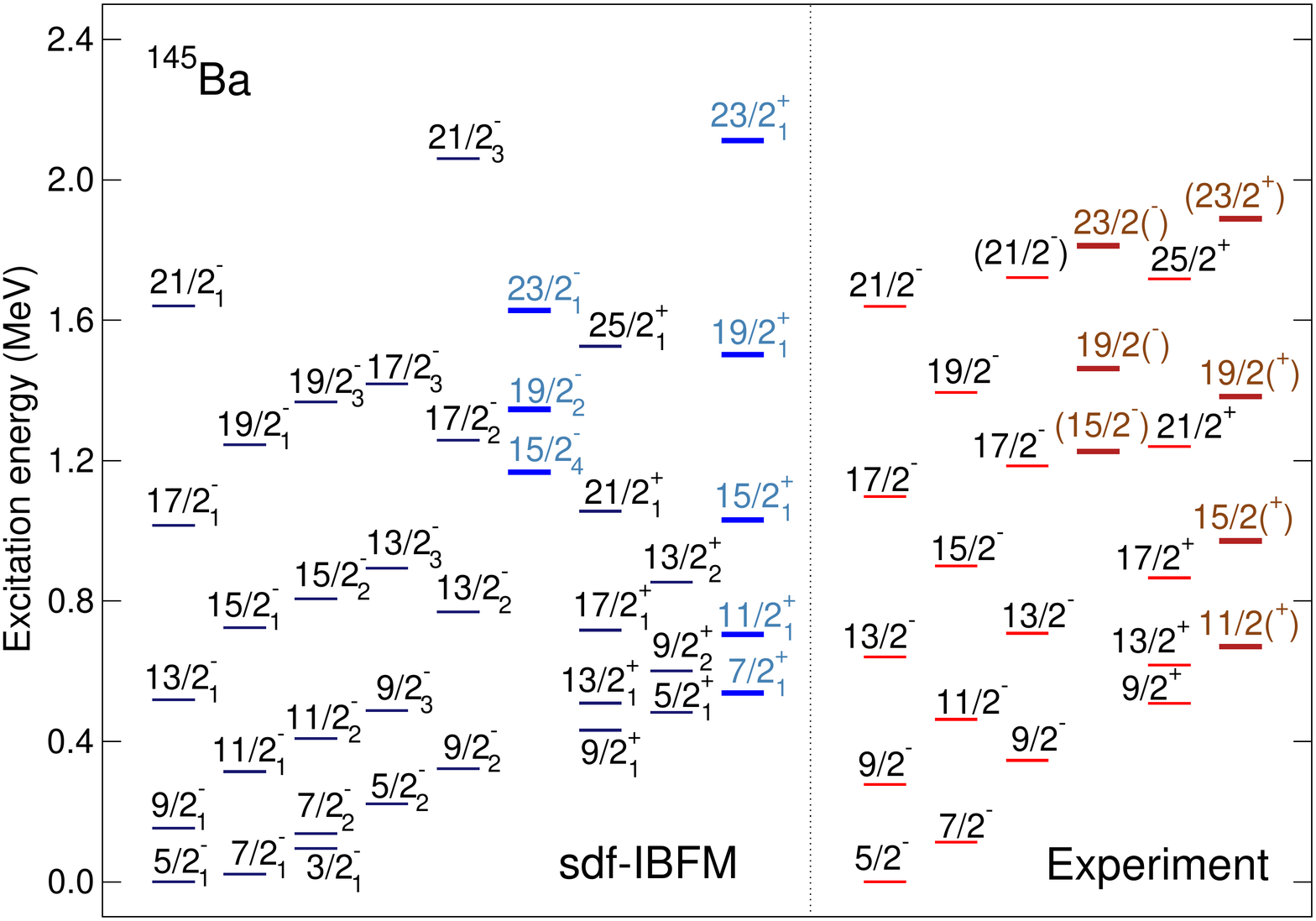}
}
\caption{Low-energy level schemes for $^{144}$Ba and $^{145}$Ba. }
\label{fig:ba}
\end{figure}

Octupole degree of freedom is expected to 
play an important role in several specific mass regions. 
Figure~\ref{fig:oct} depicts the axially-symmetric 
quadrupole $\beta_{20}$ - octupole $\beta_{30}$ 
PESs for $^{140-150}$Ba, 
where the octupole correlation is suggested to be 
particularly relevant. 
The PESs indicate pronounced octupole deformation with the 
non-zero $\beta_{30}$ value for 
$^{144-150}$Ba. 
For studying the octupole 
collective states, we included the negative-parity $f$ 
bosons with $L^\pi=3^-$ in the IBM space, in addition to the 
positive-parity $s$ ($L^\pi=0^+$) and $d$ ($L^\pi=2^+$) bosons. 
The $sdf$-IBM Hamiltonian is 
fixed by mapping the ($\beta_{20},\beta_{30}$) 
SCMF PES onto the bosonic one. 
As for the odd-A Ba, we have implemented 
the $f$-boson degrees of freedom in the IBFM 
\cite{nomura2018oct}, and 
the $sdf$-IBFM Hamiltonian has been fixed 
by a similar procedure to the one in the case of $sd$-IBFM. 
The resultant  $sdf$-IBM  
and $sdf$-IBFM spectra for even-even $^{144}$Ba 
and odd-A $^{145}$Ba 
are compared to the experimental counterparts in 
Fig.~\ref{fig:ba}. Our calculation describes very nicely 
the experimental data \cite{bucher2016} for the excitation 
spectra and $B$(E3) transition rates 
in the even-even $^{144}$Ba nucleus. 
For the odd-A $^{145}$Ba nucleus, 
the band-head energies of those bands that are empirically 
suggested \cite{rzacaurban2012} to be the octupole 
bands (shown in thick lines in Fig.~\ref{fig:ba}) are 
reproduced by the calculation 
(the corresponding theoretical bands composed of 
one-$f$-boson configuration are indicated also as 
thick lines in the figure). 
We have also predicted the $B$(E3) values to be 
approximately 20--30 W.u., for the transitions from 
the octupole to the ground state bands in odd-A Ba, 
which are comparable to the calculated 
$B$(E3$; 3^-_1\rightarrow 0^+_1)=23$ W.u. for the 
even-even neighbor.

\section{Odd-odd nuclei}

\begin{figure}[htb]
\centerline{%
\includegraphics[width=6.0cm]{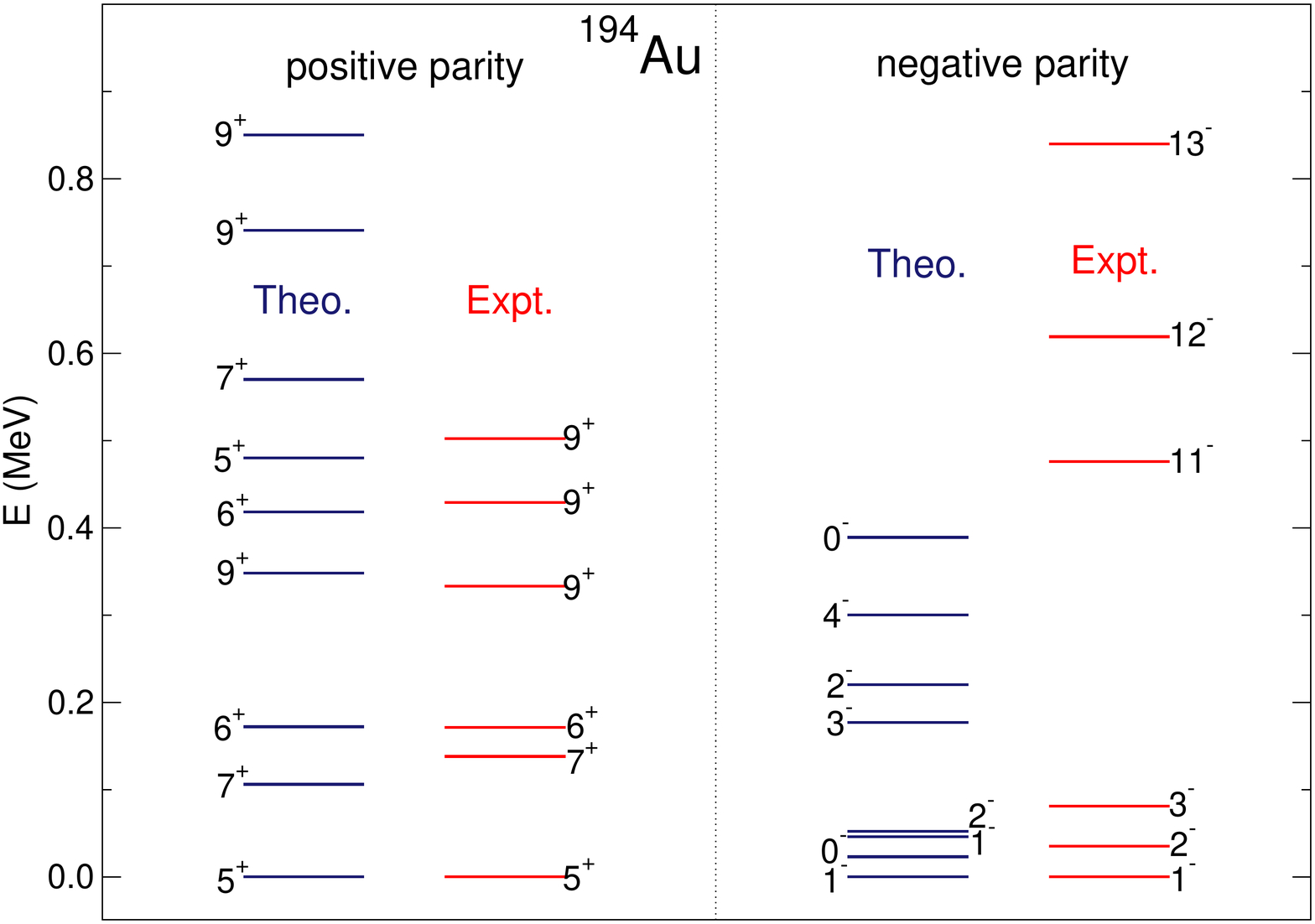}
\includegraphics[width=6.0cm]{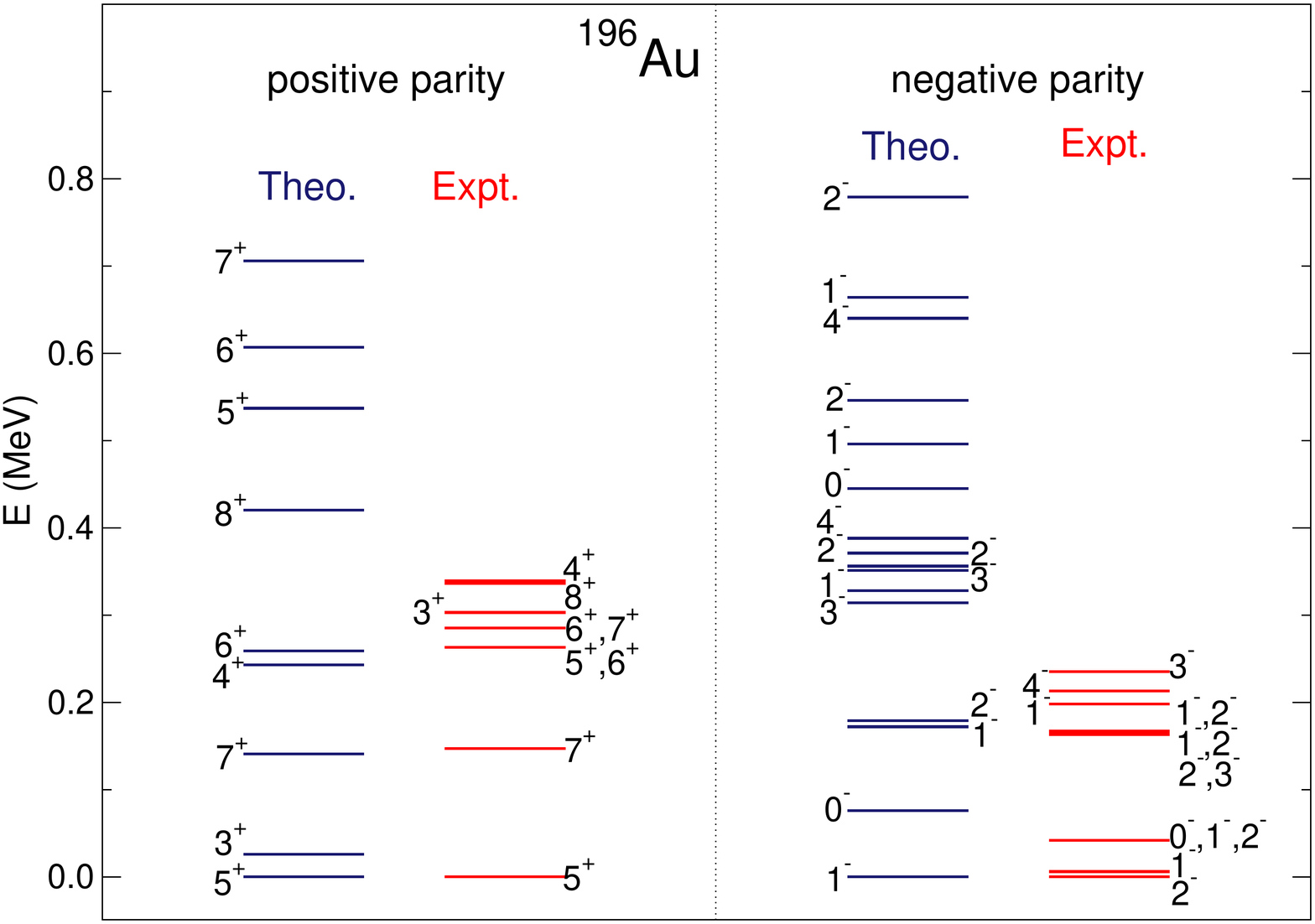}
}
\caption{Low-energy excitation spectra for the odd-odd 
nuclei $^{194,196}$Au.}
\label{fig:au}
\end{figure}

In the cases of odd-odd systems, one single neutron and 
one single proton degrees of freedom are explicitly 
considered  in the framework of 
the interacting boson-fermion-fermion model (IBFFM) \cite{IBFM}. 
The coupling constants of the interactions between 
single neutron (proton) to the boson space are 
fixed to reproduce low-energy levels of the neighboring 
odd-N (odd-Z) nucleus, using the prescription mentioned 
in Sec.~\ref{sec:model}. Furthermore, the residual 
neutron-proton interaction should be considered, 
and the parameters for the interaction are 
determined so as to reasonably reproduce the 
lowest-lying levels in the odd-odd nucleus \cite{nomura2019dodd}. 
As an example, we show in Fig.~\ref{fig:au} the IBFFM 
spectra for the odd-odd nuclei $^{194,196}$Au. 
Microscopic inputs have been provided by the 
Gogny-EDF SCMF calculation. 
The even-even core $^{194,196}$Hg 
isotopes exhibit weakly deformed oblate shapes in the 
Gogny PESs. 
The description of the excitation spectra for the 
considered odd-odd nuclei is fairly good. 
Also their electromagnetic properties have been described 
reasonably well \cite{nomura2019dodd}. 
We also mention another relevant application of the 
EDF-based IBFFM calculation of Ref.~\cite{nomura2019cs}, 
where we explored chiral band structure in a number of odd-odd Cs isotopes, 
that is mainly composed of the $(\nu h_{11/2})^{-1}\otimes\pi h_{11/2}$ 
neutron-proton pairs coupled to $\gamma$-soft even-even Xe cores. 

\section{Conclusions}

We introduced a recently developed theoretical method 
for calculating spectroscopic 
properties of those nuclei with odd nucleon numbers, that 
is based on the nuclear DFT and the particle-boson 
coupling scheme. 
The boson-core Hamiltonian, and the spherical single-particle 
energies and occupation probabilities of the odd nucleons 
are determined from fully microscopic SCMF calculations, 
whereas there are only a few free parameters that are 
fixed empirically. 
Successful applications of the employed theoretical 
method to study the shape QPT and octupole 
correlations in odd-A systems, and the structure of 
odd-odd nuclei (as well as other examples not 
covered in this contribution) indicate 
that the method is promising for 
studying spectroscopic properties of even-even, 
odd-A, and odd-odd nuclear systems in a 
systematic and computationally feasible way. 

\section*{Acknowledgements}
The results presented in this contribution are based on the works with 
D. Vretenar, T. Nik{\v s}i\'c, L. M. Robledo, and R. Rodr\'iguez-Guzm\'an,
This work is financed within the Tenure Track Pilot Programme of the 
Croatian Science Foundation and the 
\'Ecole Polytechnique F\'ed\'erale de Lausanne and the Project TTP-2018-07-3554
 Exotic Nuclear Structure and Dynamics, with funds of the Croatian-Swiss Research Programme.

\end{document}